\documentclass[preprint,showpacs,preprintnumbers,amsmath,amssymb,showkeys,nofootinbib]{revtex4}

\usepackage[cp1251]{inputenc}

\usepackage{graphicx}
\usepackage{dcolumn}
\usepackage{bm}

\usepackage{float}
\usepackage{stmaryrd} 

\usepackage{version}

\usepackage[unicode,bookmarks,bookmarksopen,bookmarksopenlevel=0,colorlinks,%
pageanchor=true,breaklinks=true,linkcolor=red,citecolor=blue,urlcolor=red,
pdfencoding = auto,  bookmarksdepth = 4]{hyperref}  

\renewcommand{\theequation}{\arabic{section}.\arabic{equation}}

\begin{document}


\title{Hodograph Method and Numerical Solution of the \\ Two Hyperbolic Quasilinear  Equations. \\ Part III. Two-Beam Reduction of the Dense \\ Soliton Gas Equations}

\author{E.\,V.~Shiryaeva}%
 \email{shir@math.sfedu.ru}
\affiliation{%
Institute of Mathematics, Mechanics and Computer Science, Southern Federal University, Rostov-on-Don, Russia
}%

\author{M.\,Yu.~Zhukov}
\email{zhuk@math.sdedu.ru}
\affiliation{%
Institute of Mathematics, Mechanics and Computer Science, Southern Federal University, Rostov-on-Don, Russia
}%

\date{\today}

\begin{abstract}

The paper presents the solutions for the two-beam reduction of the dense soliton gas equations (or Born-Infeld equation) obtained by analytical and numerical methods. The method proposed by the authors is used. This method allows to reduce the Cauchy problem for two hyperbolic quasilinear PDE's to the  Cauchy problem for ODE's. In some respect, this method is analogous to the method of characteristics for two hyperbolic equations. The method is effectively applicable in all cases when the explicit expression for the  Riemann--Green function for some linear second order PDE, resulting from the use of the hodograph method for the original equations, is known.
The numerical results for the two-beam reduction of the dense soliton gas equations, and the shallow water equations (omitting in the previous papers) are presented.
For computing we use the different initial data (periodic, wave packet).

\end{abstract}

\pacs{02.30.Jr, 02.30.Hq, 02.60.-xm, 47.15.gm}





\keywords{hodograph method,  two-beam reduction of the dense soliton gas equations, numerical method, shallow water equations}
\maketitle

\section{Introduction}\label{zhshel:sec:introd}

In previous papers \cite{Zhuk_Shir_ArXiv_2014_1,Zhuk_Shir_ArXiv_2014_2} the efficient numerical method, allowing to get solutions, including multi-valued solutions\footnote{In \cite{Zhuk_Shir_ArXiv_2014_1} the solutions of the shallow water equations describing breaking waves are presented.}, of the Cauchy problem for two hyperbolic quasilinear PDE's are presented. This method is based on the results of the paper \cite{SenashovYakhno} in which the hodograph method based on
conservation laws for two hyperbolic quasilinear PDE's is presented.

The paper \cite{SenashovYakhno} shows that the solution of the original equations can easily be written in implicit analytical form if
there is an analytical expression of the  Riemann--Green function for some linear hyperbolic equation arising as result of the hodograph transformation.
The paper \cite{Zhuk_Shir_ArXiv_2014_1} shows that one can not only write the solution in  implicit analytical form, but also construct efficient numerical method of the Cauchy problem integration. Using minor modifications of the results of paper \cite{SenashovYakhno} it is able to reduce the Cauchy problem for two quasilinear PDE's to the Cauchy problem for ODE's.
From the authors point of view, solving of the Cauchy problem for ODE's, in particular, with the help of the numerical methods,  is much easier than  solving of nonlinear transcendental equations that must be solved when there is an implicit solution of the original problem.

A key role for the proposed method plays the possibility of constructing an explicit expression  for the Riemann--Green function of the corresponding linear equation. This, of course, limits the application of the method. However, the number of the equations admitted application of this method is large enough. These include the shallow water equations (see, \emph{e.g.} \cite{RozhdestvenskiiYanenko,Whithem}), the gas dynamics equations for a polytropic gas \cite{RozhdestvenskiiYanenko,Whithem}, the two-beam reduction of the dense soliton gas equations \cite{Whithem,GenaEl} (or Born--Infeld equation), the chromatography equations  for classical isotherms \cite{RozhdestvenskiiYanenko,FerapontovTsarev_MatModel,Kuznetsov}, the isotachophoresis and zonal electrophoresis equations \cite{BabskiiZhukovYudovichRussian,ZhukovMassTransport,ZhukovNonSteadyITP,ElaevaMM,Elaeva_ZhVM}. In particular, the paper \cite{SenashovYakhno} presents a
large number of equations for which the  explicit expressions for the  Riemann--Green functions is known.
Classification of equations that allow an explicit expressions for the  Riemann--Green functions, is contained in \cite{Copson,Courant,Ibragimov} (see also \cite{Chirkunov,Chirkunov_2}).

This paper presents analytical and numerical solution of the Cauchy problem for the two-beam reduction of the dense soliton gas equations \cite{Whithem,GenaEl}.

The choice of these problem, in particular,  due to the fact that the corresponding Riemann--Green function is very simple. We emphasize that the presented results only demonstrate the method effectiveness and do not claim to any physical interpretation.
Pay attention to the fact that in some sense, the proposed method is `exact'. Its realization  does not require any approximation of the original hyperbolic PDE's, which use of the finite-difference methods, finite element method, finite volume method, the Riemann solver, etc.
Also there is no need to introduce an artificial viscosity\footnote{The effect of the grid viscosity does not occur due to the absence of approximation}. In other words, the original problem is solved without any approximation and modification. The accuracy of the solution is determined by only the accuracy of the ODE's numerical solution method.

The paper is organized as follows. In Sec.~II the Cauchy problem for the two-beam reduction of the dense soliton gas equations is formulated. Here we construct the densities and fluxes of some conservation laws (Sec.~\ref{zhshArXiv:sec:4.A}), the implicit solution of the problem (Sec.~\ref{zhshArXiv:sec:4.B}),
the solution on the isochrone (Sec.~\ref{zhshArXiv:sec:4.C}). In Sec.~\ref{zhshArXiv:sec:4.D} we show the impossibility of the breaking solution and investigate the properties of the discontinuity solutions. The numerical results are contained in Sec.~\ref{zhshArXiv:sec:4.E}. In Sec.~\ref{zhshArXiv:sec:5} we present the some numerical results for shallow water equations omitted in previous paper \cite{Zhuk_Shir_ArXiv_2014_1}. Appendix~\ref{zhshArXiv:sec:2} gives the short description of the numerical methods (more detail see in \cite{Zhuk_Shir_ArXiv_2014_1,Zhuk_Shir_ArXiv_2014_2}).

\setcounter{equation}{0}

\section{Two-beam reduction of the dense soliton gas equations}\label{zhshArXiv:sec:4}

To demonstrate the effectiveness of the hodograph method based on the conservation laws we  consider the equation, the so-called two-beam reduction of the
 dense soliton gas equations \cite{GenaEl} (notations are changed)
\begin{equation}\label{zhshArXiv:eq:4.01}
u^1_t+(u^1 R^1)_x=0, \quad u^2_t+(u^2 R^2)_x=0,
\end{equation}
\begin{equation}\label{zhshArXiv:eq:4.02}
R^1=4\alpha\frac{1-\kappa(u^1-u^2)}{1-\kappa(u^1+u^2)}, \quad
R^2=-4\alpha\frac{1+\kappa(u^1-u^2)}{1-\kappa(u^1+u^2)}, \quad
R^1 \ne R^2,
\end{equation}
where $\alpha$, $\kappa$ are the parameters.

Note that these equations after some transformations are also well known as the Born--Infeld equation (see, e.g., \cite{Whithem}), which is investigated enough detailed in \cite{Menshikh01,Menshikh02,Menshikh03,Menshikh04}.

The equation (\ref{zhshArXiv:eq:4.01}) can be rewritten in the Riemann invariants  $R^1$, $R^2$
\begin{equation}\label{zhshArXiv:eq:4.03}
R^1_t+R^2 R^1_x=0, \quad R^2_t+R^1 R^2_x=0,
\end{equation}
\begin{equation}\label{zhshArXiv:eq:4.04}
\lambda^1=R^2,  \quad \lambda^2=R^1.
\end{equation}
The connection of the  Riemann invariants with the original variables given by (\ref{zhshArXiv:eq:4.02}) has the following form
\begin{equation}\label{zhshArXiv:eq:4.05}
u^1=\frac{R^2+4\alpha}{\kappa(R^2-R^1)}, \quad
u^2=\frac{R^1-4\alpha}{\kappa(R^1-R^2)}.
\end{equation}

The original notations of the paper \cite{GenaEl} have the following form
\begin{equation}\label{zhshArXiv:eq:4.06}
u^1=\rho_1, \quad u^2=\rho_2, \quad R^1=s_1, \quad R^2=s_2,
\end{equation}
where $\rho_1$, $\rho_2$ are the densities and $s_1$, $s_2$ are the velocities.


\subsection{Densities and fluxes of the conservation laws.}\label{zhshArXiv:sec:4.A}

To obtain the density $\varphi^t$ and flux $\psi^t$ of a conservation law
\begin{equation}\label{zhshArXiv:eq:4.07}
\varphi^t_t+\psi^t_x=0,
\end{equation}
which satisfy the conditions (\ref{zhshArXiv:eq:2.08})
\begin{equation}\label{zhshArXiv:eq:4.08}
(\psi^t - \lambda^1 \varphi^t)\bigr|_{{R^1=r^1}}=1, \quad (\psi^t - \lambda^2 \varphi^t)\bigr|_{{R^2=r^2}}=-1,
\end{equation}
we use the natural conservation laws (\ref{zhshArXiv:eq:4.01}).

We represent functions $\varphi^t$, $\psi^t$ as a linear combination of the functions $u^1$ and $u^2$
\begin{equation}\label{zhshArXiv:eq:4.09}
\varphi^t=\beta^1 u^1 + \beta^2 u^2 + \beta^0, \quad
\psi^t=\beta^1 R^1 u^1 + \beta^2 R^2 u^2 + \beta.
\end{equation}
Here, $\beta^1$, $\beta^2$, $\beta^0$, $\beta$ are arbitrary functions depended on  $r^1$, $r^2$.

Substitution (\ref{zhshArXiv:eq:4.09}) in (\ref{zhshArXiv:eq:4.08}) and identical satisfying of (\ref{zhshArXiv:eq:4.08}) gives
\begin{equation}\label{zhshArXiv:eq:4.10}
\beta^1=-\frac{\kappa}{4\alpha}, \quad
\beta^2=-\frac{\kappa}{4\alpha}, \quad
\beta^0=\frac{\kappa}{4\alpha}, \quad
\beta=0.
\end{equation}
Using (\ref{zhshArXiv:eq:4.10}) and  (\ref{zhshArXiv:eq:4.09}) we get (see also \cite{SenashovYakhno})
\begin{equation}\label{zhshArXiv:eq:4.11}
\varphi^t=\frac{2}{R^1-R^2}, \quad
\psi^t=\frac{R^1+R^2}{R^1-R^2}.
\end{equation}

Another conservation law
\begin{equation}\label{zhshArXiv:eq:4.12}
\varphi^x_t+\psi^x_x=0,
\end{equation}
which satisfies the conditions (\ref{zhshArXiv:eq:2.09})
\begin{equation}\label{zhshArXiv:eq:4.13}
\left(
\frac{\psi^x}{\lambda^1} - \varphi^x
\right)_{{R^1=r^1}}=1,
\quad \left(
\frac{\psi^x}{\lambda^2} - \varphi^x
\right)_{{R^2=r^2}}=-1
\end{equation}
can be constructed by analogous.

We assume that $\varphi^x$, $\psi^x$ are the linear combination
\begin{equation}\label{zhshArXiv:eq:4.14}
\varphi^x=\beta^1 u^1 + \beta^2 u^2 + \beta^0, \quad
\psi^x=\beta^1 R^1 u^1 + \beta^2 R^2 u^2 + \beta.
\end{equation}
Identical satisfying of the conditions (\ref{zhshArXiv:eq:4.13}) gives  $\beta^1$, $\beta^2$, $\beta^0$, $\beta$
\begin{equation}\label{zhshArXiv:eq:4.15}
\beta^1=-\kappa, \quad
\beta^2=\kappa, \quad
\beta^0=0, \quad
\beta=-4\alpha,
\end{equation}
and functions $\varphi^x$, $\psi^x$ (see also \cite{SenashovYakhno})
\begin{equation}\label{zhshArXiv:eq:4.16}
\varphi^x=\frac{R^1+R^2}{R^1-R^2}, \quad
\psi^x=\frac{2R^1R^2}{R^1-R^2}.
\end{equation}

Note that functions $\varphi^t$, $\varphi^x$, $\psi^t$, $\psi^x$  depend only on the variables $R^1$, $R^2$ and do not depend on the variables $r^1$, $r^2$.


\subsection{Implicit solution of the problem}\label{zhshArXiv:sec:4.B}

Taking into account the simple form of the functions $\varphi^t$, $\varphi^x$, $\psi^t$, $\psi^x$ we present the solution
of the Cauchy problem for equations (\ref{zhshArXiv:eq:4.03}), (\ref{zhshArXiv:eq:4.04}) with initial data given on arbitrary curve.

We assume that initial data for the equations (\ref{zhshArXiv:eq:4.03}), (\ref{zhshArXiv:eq:4.04}) are given for some line $\Gamma$ (not a characteristic)
\begin{equation}\label{zhshArXiv:eq:4.17}
\Gamma=\{(x,t)\,:\, x=x(\tau), \quad t=t(\tau)\},
\end{equation}
\begin{equation}\label{zhshArXiv:eq:4.18}
R^1\bigr|_\Gamma=R^1_0(\tau), \quad R^2\bigr|_\Gamma=R^2_0(\tau).
\end{equation}
Here, $R^1_0(\tau)$, $R^2_0(\tau)$ are given functions, $\tau$ is the parameter.

Using the hodograph method based on conservation laws (see \cite{SenashovYakhno}) we get
\begin{equation}\label{zhshArXiv:eq:4.19}
2t=t(a)+t(b)-\int\limits_{\Gamma}(\psi^t dt - \varphi^t dx),
\end{equation}
\begin{equation}\label{zhshArXiv:eq:4.20}
2x=x(a)+x(b)-\int\limits_{\Gamma}(\psi^x dt - \varphi^x dx),
\end{equation}
where the functions $\varphi^t$, $\psi^t$ are determined by relations  (\ref{zhshArXiv:eq:4.11}), and the functions $\varphi^x$, $\psi^x$ are determined by relations (\ref{zhshArXiv:eq:4.16}).

We restrict the investigation by the easiest and most natural situation, when the initial data is given at  $t=t_0$. In this case, the contour $\Gamma$ is an interval of axis $t=t_0$, and $\tau=x$ (for $x$ more convenient is to keep the  previous notation $\tau$). The conditions (\ref{zhshArXiv:eq:4.18}) take the form
\begin{equation}\label{zhshArXiv:eq:4.21}
R^1\bigr|_{t=t_0}=R^1_0(\tau), \quad R^2\bigr|_{t=t_0}=R^2_0(\tau).
\end{equation}
Then
\begin{equation}\label{zhshArXiv:eq:4.22}
t=t_0+\frac12\int\limits_{a}^{b}\varphi^t\,d\tau, \quad x=\frac{a+b}{2}+\frac12\int\limits_{a}^{b}\varphi^x \,d\tau.
\end{equation}

We introduce the notations
\begin{equation}\label{zhshArXiv:eq:4.23}
F(a,b)=\frac12\int\limits_{a}^{b}\varphi^t\,d\tau=
\int\limits_{a}^{b}\frac{1}{R^1_0(\tau)-R^2_0(\tau)}\,d\tau,
\end{equation}
\begin{equation}\label{zhshArXiv:eq:4.24}
G(a,b)=\frac12\int\limits_{a}^{b}\varphi^x\,d\tau=
\frac12\int\limits_{a}^{b}\frac{R^1_0(\tau)+R^2_0(\tau)}{R^1_0(\tau)-R^2_0(\tau)}\,d\tau,
\end{equation}
where $F(a,b)$, $G(a,b)$ are completely determined by the initial data, and they depend only on the parameters $a$, $b$.

The implicit solution of the problem (\ref{zhshArXiv:eq:4.03}), (\ref{zhshArXiv:eq:4.04}), (\ref{zhshArXiv:eq:4.21}) takes the form
\begin{equation}\label{zhshArXiv:eq:4.25}
t=t(a,b) \equiv t_0+F(a,b), \quad x=x(a,b) \equiv \frac{a+b}{2}+G(a,b).
\end{equation}
\begin{equation}\label{zhshArXiv:eq:4.26}
R^1(x,t)=r^1 \equiv R^1_0(b), \quad R^2(x,t)=r^2 \equiv R^2_0(a).
\end{equation}

Also we present the implicit solutions of the original problem (\ref{zhshArXiv:eq:4.01}), (\ref{zhshArXiv:eq:4.02}).
Taking into account (\ref{zhshArXiv:eq:4.02}), (\ref{zhshArXiv:eq:4.05}) we obtain
\begin{equation}\label{zhshArXiv:eq:4.27}
R^1-R^2=\frac{8\alpha}{1-\kappa(u^1+u^2)}, \quad
R^1+R^2=\frac{8\alpha\kappa(u^2-u^1)}{1-\kappa(u^1+u^2)},
\end{equation}
\begin{equation}\label{zhshArXiv:eq:4.28}
F(a,b)=\int\limits_{a}^{b}
\frac{1-\kappa(u^1_0(\tau)+u^2_0(\tau))}{8\alpha}\,d\tau,
\end{equation}
\begin{equation}\label{zhshArXiv:eq:4.29}
G(a,b)=
\int\limits_{a}^{b}
\frac{\kappa(u^2_0(\tau)-u^1_0(\tau))}{2}\,d\tau,
\end{equation}
\begin{equation}\label{zhshArXiv:eq:4.30}
u^1(x,t)=\frac{u^1_0(a)(1-\kappa(u^1_0(b)+u^2_0(b)))}{1-\kappa(u^1_0(b)+u^2_0(a))-\kappa^2(u^1_0(a) u^2_0(b)-u^1_0(b) u^2_0(a))},
\end{equation}
\begin{equation}\label{zhshArXiv:eq:4.31}
u^2(x,t)=\frac{u^2_0(b)(1-\kappa(u^1_0(a)+u^2_0(a)))}{1-\kappa(u^1_0(b)+u^2_0(a))-\kappa^2(u^1_0(a) u^2_0(b)-u^1_0(b) u^2_0(a))}.
\end{equation}
Here, $u^1_0$, $u^2_0$ are the initial data at $t=t_0$.


\subsection{The solution on isochrone}\label{zhshArXiv:sec:4.C}

We describe the solving of the Cauchy problem for the isochrones, i.e. on line level of function $t(a,b)$, which is determined by  (\ref{zhshArXiv:eq:4.23})--(\ref{zhshArXiv:eq:4.25}). We recall that the function $x(a,b)$ and hence $G(a,b)$ are not required.

Calculating the derivative of $t_a(a,b)$, $t_b(a,b)$, i.\,e. the right parts of the differential equations (\ref{zhshArXiv:eq:2.18}), we get with the help of (\ref{zhshArXiv:eq:4.23})
\begin{equation}\label{zhshArXiv:eq:4.32}
t_a=t_a(a,b)=-f(a), \quad t_b=t_b(a,b)=f(b),
\end{equation}
\begin{equation}\label{zhshArXiv:eq:4.33}
f(\tau)=\frac{1}{R^1_0(\tau)-R^2_0(\tau)}.
\end{equation}

We assume that isochrone is given by the parameters $a_*$, $b_*$
\begin{equation}\label{zhshArXiv:eq:4.34}
t_*=t(a_*, b_*).
\end{equation}

To determine the coordinates $X_*=x(a_*,b_*)$, corresponding to  the parameter $\tau=0$ we have the Cauchy problem (\ref{zhshArXiv:eq:2.21}), (\ref{zhshArXiv:eq:2.22}), which
can be written with the help of (\ref{zhshArXiv:eq:4.03}), (\ref{zhshArXiv:eq:4.32}), (\ref{zhshArXiv:eq:4.33}) in the following form
\begin{equation}\label{zhshArXiv:eq:4.35}
\frac{dY(b)}{db}=\frac{R^1_0(b)}{R^1_0(b)-R^2_0(b)}, \quad Y(a_*)=a_*.
\end{equation}
Integrating from $a_*$ to $b_*$ we have
\begin{equation}\label{zhshArXiv:eq:4.36}
X_*=Y(b_*)=a_*+\int\limits_{a_*}^{b_*}\frac{R^1_0(b)\, db}{R^1_0(b)-R^2_0(b)} .
\end{equation}

To determine the functions $a(\tau)$, $b(\tau)$ and $x=X(\tau)$  we get the Cauchy problem (\ref{zhshArXiv:eq:2.18}), (\ref{zhshArXiv:eq:2.20}), using again
(\ref{zhshArXiv:eq:4.03}), (\ref{zhshArXiv:eq:4.32}), (\ref{zhshArXiv:eq:4.33})
\begin{equation}\label{zhshArXiv:eq:4.37}
\frac{da}{d\tau}=\frac{1}{R^2_0(b)-R^1_0(b)}, \quad
\frac{db}{d\tau}=\frac{1}{R^2_0(a)-R^1_0(a)},
\end{equation}
\begin{equation}\label{zhshArXiv:eq:4.38}
\frac{dX}{d\tau}=\frac{R^2_0(a)-R^1_0(b)}{(R^1_0(a)-R^2_0(a))(R^1_0(b)-R^2_0(b))}.
\end{equation}
\begin{equation}\label{zhshArXiv:eq:4.39}
a\bigr|_{\tau=0}=a_*, \quad b\bigr|_{\tau=0}=b_*, \quad
X\bigr|_{\tau=0}=X_*.
\end{equation}
Integrating the (\ref{zhshArXiv:eq:4.37})--(\ref{zhshArXiv:eq:4.39})
we obtain the solutions on isochrone
\begin{equation}\label{zhshArXiv:eq:4.40}
R^1(x,t_*)=R^1_0(b(\tau)), \quad R^2(x,t_*)=R^2_0(a(\tau)), \quad
x=X(\tau).
\end{equation}


\subsection{The impossibility of profile breaking. Discontinuous solutions}\label{zhshArXiv:sec:4.D}

Before further investigation of the problem we  note that the parameters $\alpha$ and $\kappa$ can be excluded from equations with the help of the substitutions
\begin{equation}\label{zhshArXiv:eq:4.41}
t \rightarrow \frac{t}{4\alpha}, \quad \kappa u^i \rightarrow u^i.
\end{equation}
Further, we just assume
\begin{equation}\label{zhshArXiv:eq:4.42}
4\alpha=1, \quad \kappa=1.
\end{equation}

One of the breaking solution conditions at some time $t$ (i. e. the formation of the multi-valued solutions) is the tending to infinity of the derivatives  $R^1_x(x,t)$, $R^2_x(x,t)$. For example, calculating $R^1_x(x,t)$ we get
\begin{equation}\label{zhshArXiv:eq:4.43}
R^1_x(x,t)=\partial_x R^1_0(x)=r^1_b(b)b_x.
\end{equation}
Differentiating $t=t(a,b)$, $x=x(a,b)$ with respect to $x$ we have
\begin{equation}\label{zhshArXiv:eq:4.44}
x_a a_x+x_b b_x=1, \quad t_a a_x+t_b b_x=0.
\end{equation}
Then
\begin{equation}\label{zhshArXiv:eq:4.45}
a_x=\frac{t_b}{\Delta}, \quad b_x=\frac{-t_b}{\Delta}, \quad \Delta=x_a t_b - x_b t_a.
\end{equation}
Hence,
\begin{equation}\label{zhshArXiv:eq:4.46}
R^1_x(x,t)=\frac{-r^1_b(b)t_a}{\Delta}.
\end{equation}

Obviously, the breaking solution condition  is $R^1_x(x,t)=\infty$ or $\Delta=0$. Taking into account (\ref{zhshArXiv:eq:4.23})--(\ref{zhshArXiv:eq:4.26}) we calculate the dervaties and get
\begin{equation}\label{zhshArXiv:eq:4.47}
R^1_0(b)=R^2_0(a).
\end{equation}
This equality is impossible because it means that for some point $(x,t)$ we have relation
\begin{equation}\label{zhshArXiv:eq:4.48}
R^1(x,t)=R^2(x,t),
\end{equation}
which contradicts to the condition (\ref{zhshArXiv:eq:4.02}) (at $4\alpha=1$ and $\kappa=1$)
\begin{equation}\label{zhshArXiv:eq:4.49}
R^1-R^2=\frac{2}{1-u^1+u^2} \ne 0.
\end{equation}

The results obtained indicate that the breaking profile of the function $R^1(x,t)$, $R^2(x,t)$ is impossible. In other words
$R^1(x,t)$, $R^2(x,t)$ are the one-valued functions. We recall also that it is impossible to construct a self-similar solution, since the system (\ref{zhshArXiv:eq:4.03}) is the degeneracy system
\begin{equation}\label{zhshArXiv:eq:4.50}
\lambda^1_{R^1}=0, \quad \lambda^2_{R^2}=0.
\end{equation}

It means that discontinuities of solutions can be set only at the initial moment (can not occur when initial data are smooth). This discontinuity solution is so called contact discontinuity which can move along characteristics only.

The Rankine–Hugoniot conditions for the conservative system (\ref{zhshArXiv:eq:4.01}), after a change of variables (\ref{zhshArXiv:eq:4.05}), are written in the form
\begin{equation}\label{zhshArXiv:eq:4.51}
D
\left\llbracket \frac{2}{R^1-R^2} \right\rrbracket =
\left\llbracket \frac{R^1+R^2}{R^1-R^2} \right\rrbracket,
\quad
D
\left\llbracket \frac{R^1+R^2}{R^1-R^2} \right\rrbracket =
\left\llbracket \frac{2R^1R^2}{R^1-R^2} \right\rrbracket,
\end{equation}
where $\llbracket \, . \, \rrbracket$ is the jump across discontinuity, $D$ is the discontinuity velocity.

It is easy to show that there are only the following solutions of system (\ref{zhshArXiv:eq:4.51})
\begin{equation}\label{zhshArXiv:eq:4.52}
D=R^2, \quad \left\llbracket R^1 \right\rrbracket \ne 0, \quad
\quad \left\llbracket R^2 \right\rrbracket = 0
\end{equation}
or
\begin{equation}\label{zhshArXiv:eq:4.53}
D=R^1, \quad \left\llbracket R^1 \right\rrbracket = 0, \quad
\quad \left\llbracket R^2 \right\rrbracket \ne  0.
\end{equation}

In particular, the simultaneous discontinuities of the Riemann invariants (\emph{i.e.}, $\llbracket R^1 \rrbracket \ne 0$, $\llbracket R^2 \rrbracket \ne 0$) are possible either at the initial moment of time (the Riemann problem), or at intersections in the process of its motion. For example, the moving discontinuities of the Riemann invariants can intersect in some point, and then pass through each other without changing its velocities. Of course, the magnitude of the jumps of discontinuities $\llbracket R^i \rrbracket$ in the process of evolution can change its values. To avoid misunderstandings, note that the discontinuities of densities $\llbracket u^1 \rrbracket$, $\llbracket u^2 \rrbracket $ can exist simultaneously.


\subsection{Numerical results}\label{zhshArXiv:sec:4.E}

We demonstrate two examples of the initial density distribution evolution. In the first example, the initial distribution of density is periodic in space
\begin{equation}\label{zhshArXiv:eq:4.54}
u^1_0=0.2(1+0.1\cos x), \quad u^2_0=0.3(1+0.2\sin x), \quad 4\lambda=1, \quad \kappa=1.
\end{equation}

\begin{figure}[H]
\centering
\includegraphics[scale=1.00]{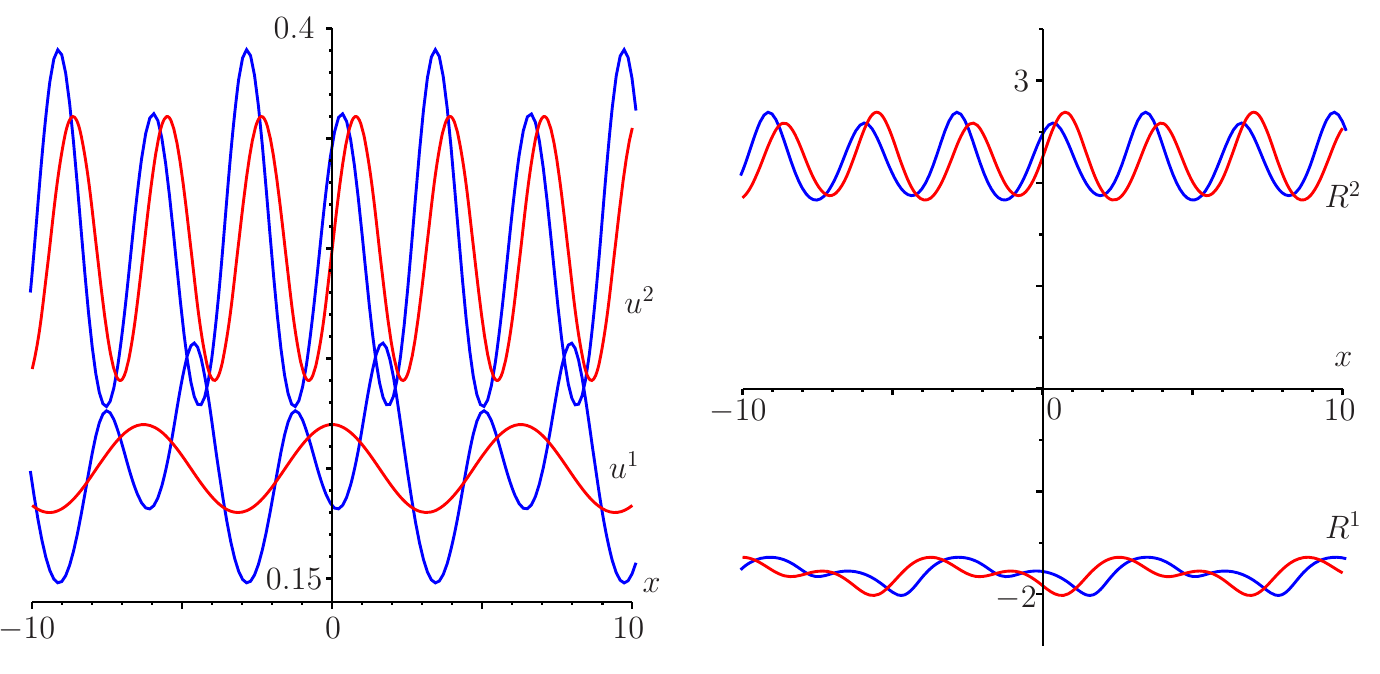}
\caption[Solitons gas, periodic solution]{
$a_*=-16.89$, $b_*=18.90$, $\delta \tau=0.5 $, $x(a_*,b_*)= 2.778$, $t(a_*,b_*)=8.982$
}
\label{fig12.5.1}
\end{figure}

On Fig.~\ref{fig12.5.1} the distribution of the densities and the Riemann invariants at time $t=8.982$ is shown. The  red lines correspond to the initial distribution.

The second example demonstrates the Riemann problem solutions. We solve the general Riemann problem when
the initial  discontinuities is not piecewise constant.
\begin{equation}\label{zhshArXiv:eq:4.55}
u^1_0=(0.2-0.1h(x+2))(1+0.1\cos 2x), \quad
4\lambda=1, \quad \kappa=1
\end{equation}
\begin{equation*}
u^2_0=(0.3+0.1h(x-2))(1+0.1\sin 3x),
\end{equation*}
where $h(\tau)$ is the Heaviside step function.

\begin{figure}[H]
\centering
\includegraphics[scale=1.00]{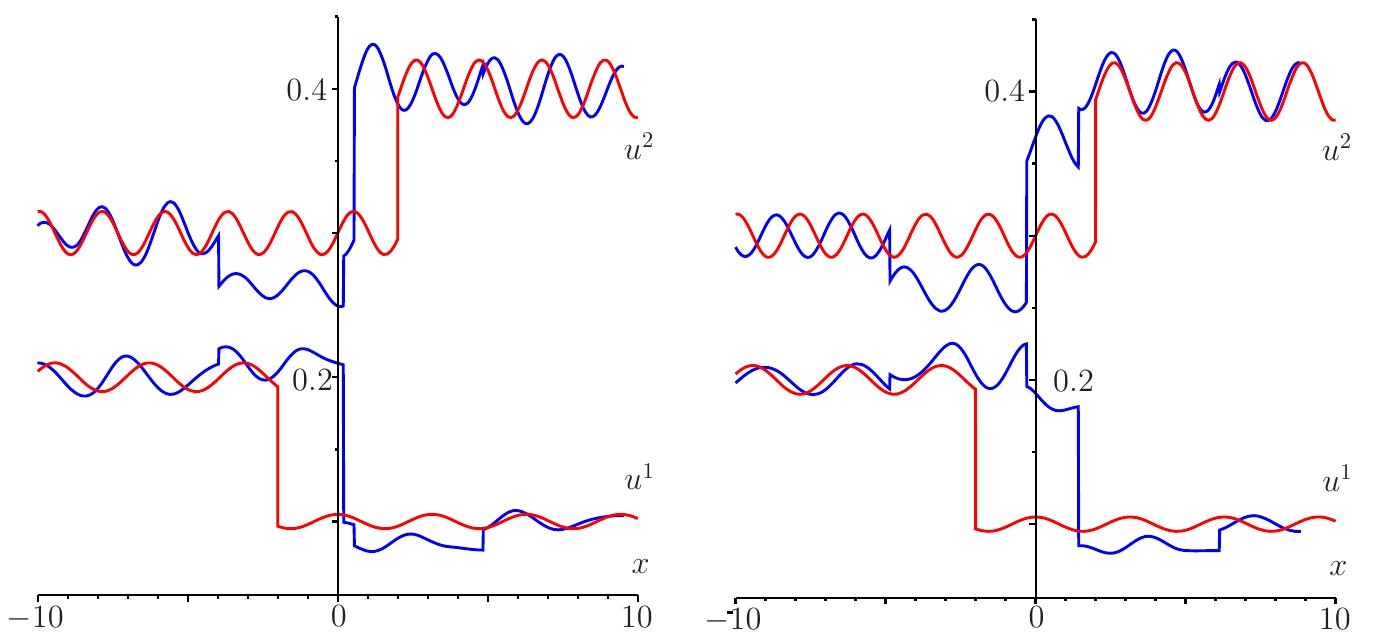}
\caption[Solitons gas, the Riemann problem]{
(left)  $a_*=-2.89$, $b_*=0.90$, $\delta \tau=0.05 $, $x(a_*,b_*)= -0.661$, $t(a_*,b_*)=1.091$;
(right) $a_*=-4.89$, $b_*=0.90$, $\delta \tau=0.05 $, $x(a_*,b_*)= -1.561$, $t(a_*,b_*)=1.590$
}
\label{fig12.5.2}
\end{figure}

On Fig.~\ref{fig12.5.2} the distribution of the densities at time $t=1.091$, $t=1.590$  are shown. The red lines correspond to the initial distribution.

\begin{figure}[H]
\centering
\includegraphics[scale=1.00]{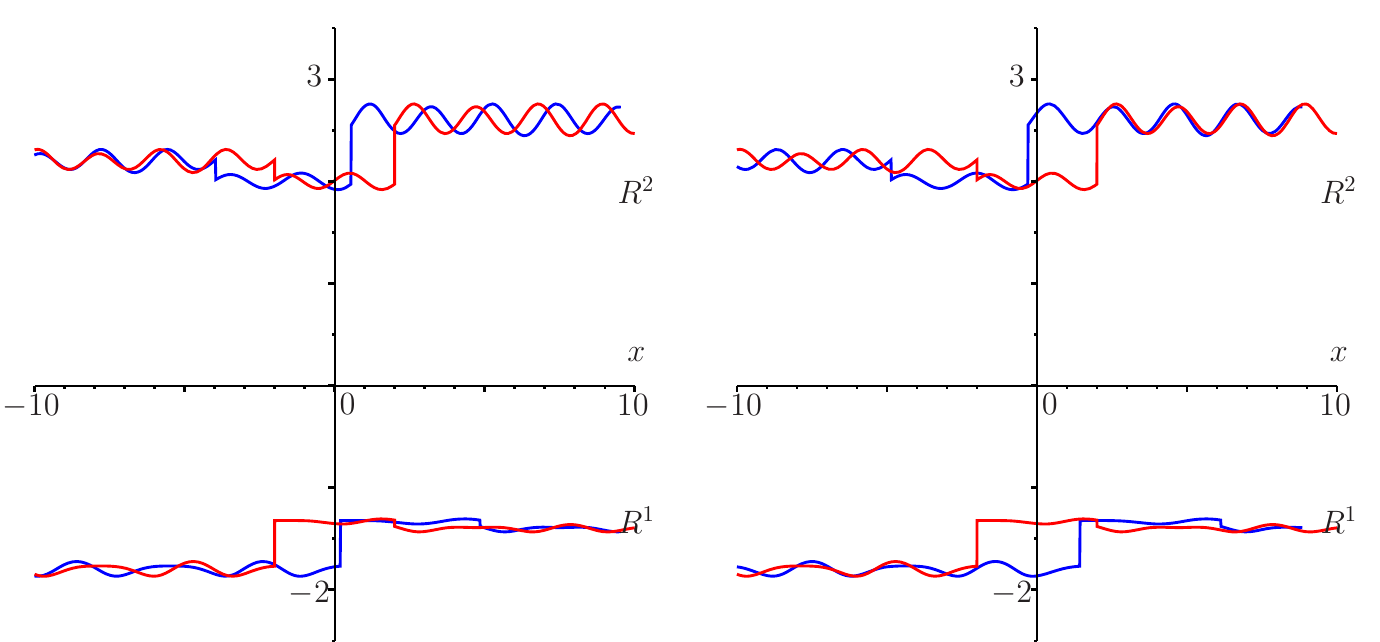}
\caption[Solitons gas, the Riemann problem]{
(left)  $a_*=-2.89$, $b_*=0.90$, $\delta \tau=0.05 $, $x(a_*,b_*)= -0.661$, $t(a_*,b_*)=1.091$;
(right) $a_*=-4.89$, $b_*=0.90$, $\delta \tau=0.05 $, $x(a_*,b_*)= -1.561$, $t(a_*,b_*)=1.590$
}
\label{fig12.5.3}
\end{figure}

On Fig.~\ref{fig12.5.3} the distribution of the Riemann invariant at time $t=1.091$, $t=1.590$  are shown. The red lines correspond to the initial distribution.

On Figs.~\ref{fig12.5.2},~\ref{fig12.5.3} the evolution of the discontinuities is well visible. As already mentioned, the simultaneous discontinuities of the invariants $R^1$, $R^2$ exist only at their the interaction. On the contrary, the discontinuities of the densities $u^1$, $u^2$ can be simultaneous.

\setcounter{equation}{0}

\section{The shallow water equations}\label{zhshArXiv:sec:5}

In this Section we present the numerical results for the shallow water equations omitted in previous paper \cite{Zhuk_Shir_ArXiv_2014_1}.

The classic version of the shallow water equations without taking into account the slope of the bottom has the form (see for example \cite{RozhdestvenskiiYanenko,Whithem})
\begin{equation}\label{zhshArXiv:eq:5.01}
h_t+(hv)_x=0, \quad v_t+\left(\frac12v^2+h\right)_x=0,
\end{equation}
where $h>0$ is the elevation of the free surface, $v$ is the velocity.

We rewrite the equations in the form
\begin{equation}\label{zhshArXiv:eq:5.02}
u^1_t+(u^1 u^2)_x=0, \quad u^2_t+\left(\frac12 u^2 u^2+u^1\right)_x=0,
\quad h=u^1, \quad v=u^2.
\end{equation}


\subsection{Interactions of the `solitons'}\label{zhshArXiv:sec:5.A}

The initial distribution
\begin{equation}\label{zhshArXiv:eq:5.03}
u^1_0=1+0.2e^{-(x+3)^2}+0.2e^{-(x-3)^2}, \quad
u^2_0=0.2e^{-(x+3)^2}-0.2e^{-(x-3)^2}
\end{equation}
simulates the interaction of two `solitons'.
The initial perturbations
of the free surface and velocity are given in the form of Gaussian distributions. The right perturbation moves to the left, and the left perturbation moves to the right.

\begin{figure}[H]
\centering
\includegraphics[scale=1.00]{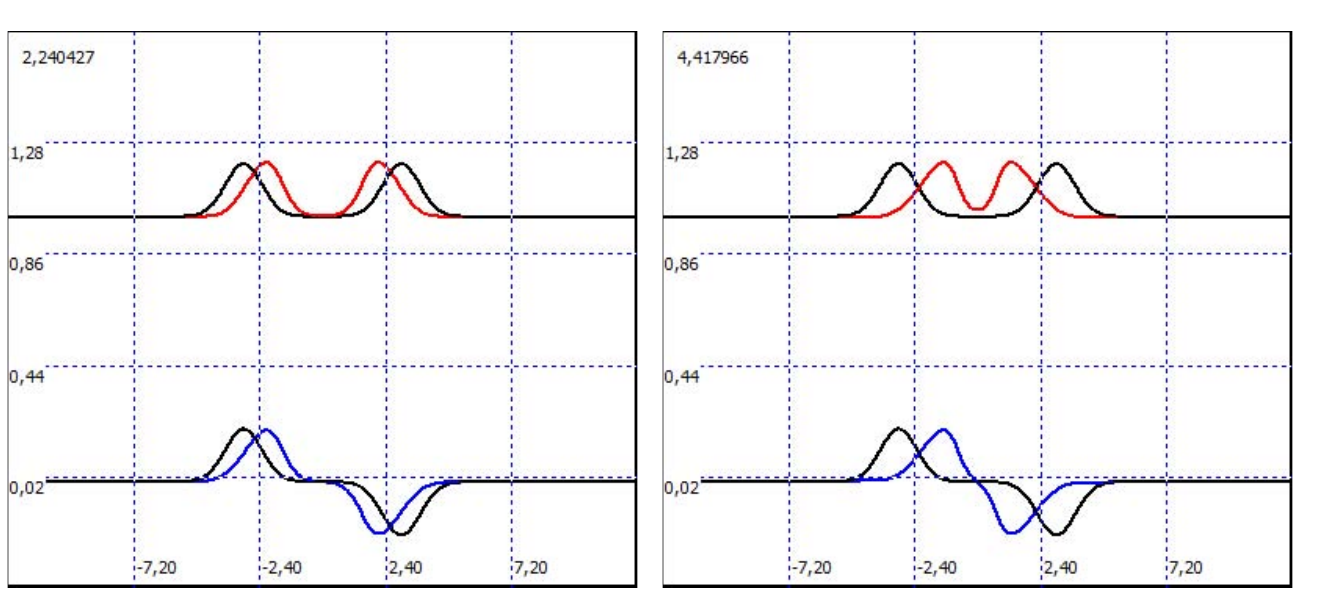}
\caption[Shallow water, waves interactions, $h=u^1$, $v=u^2$]{$h=u^1$ (red), $v=u^2$ (blue),  the initial distributions (black)}
\label{fig13.3.1}
\end{figure}

\begin{figure}[H]
\centering
\includegraphics[scale=1.00]{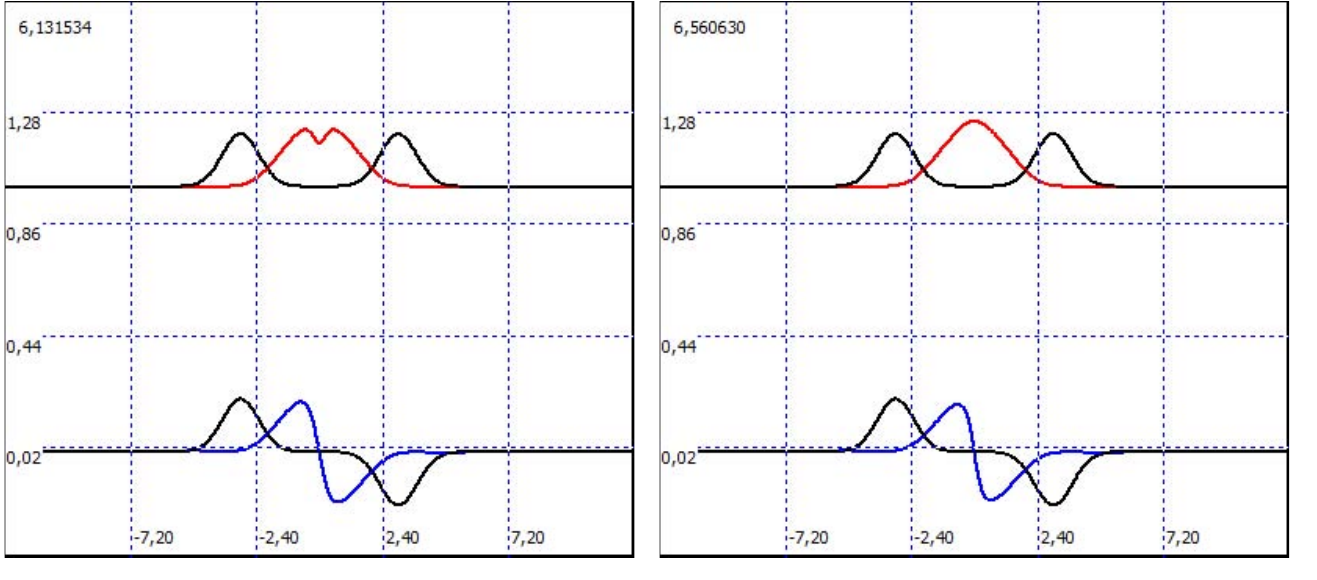}
\caption[Shallow water, waves interactions, $h=u^1$, $v=u^2$]{$h=u^1$ (red), $v=u^2$ (blue),  the initial distributions (black)}
\label{fig13.3.2}
\end{figure}

\begin{figure}[H]
\centering
\includegraphics[scale=1.00]{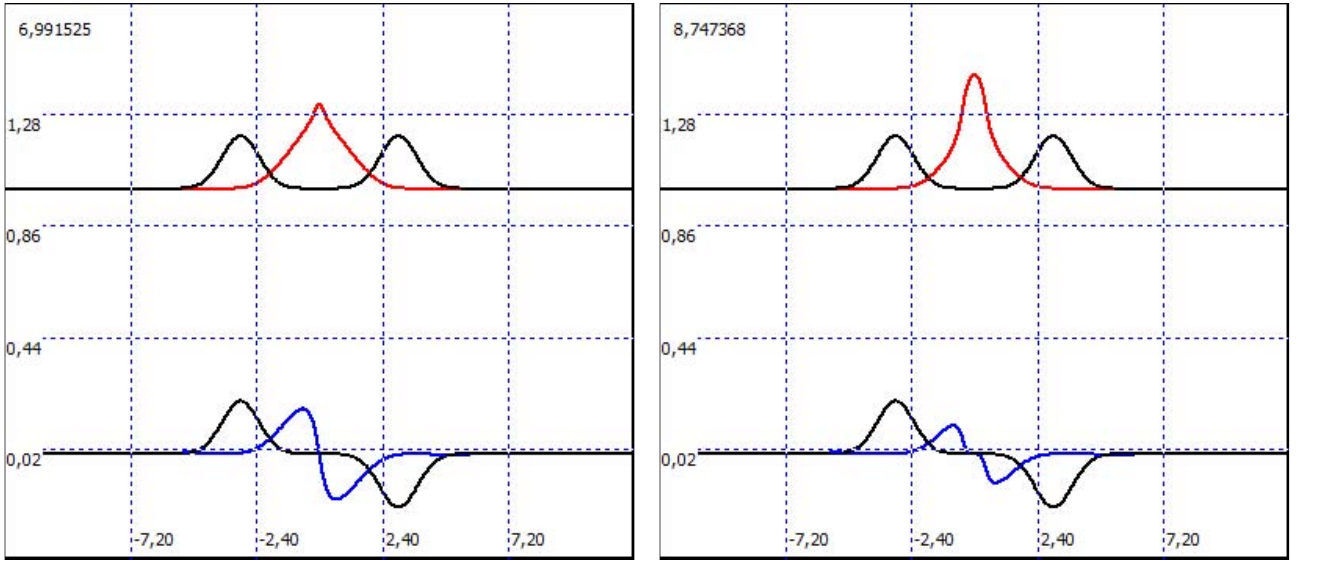}
\caption[Shallow water, waves interactions, $h=u^1$, $v=u^2$]{$h=u^1$ (red), $v=u^2$ (blue),  the initial distributions (black)}
\label{fig13.3.3}
\end{figure}

\begin{figure}[H]
\centering
\includegraphics[scale=1.00]{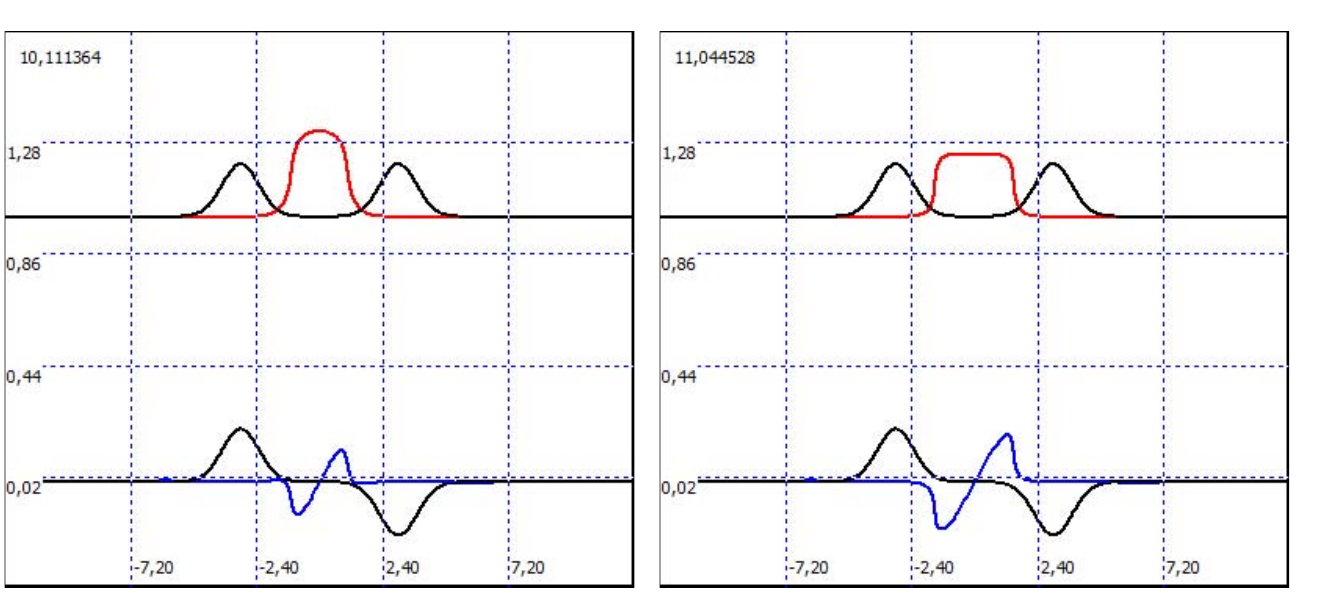}
\caption[Shallow water, waves interactions, $h=u^1$, $v=u^2$]{$h=u^1$ (red), $v=u^2$ (blue),  the initial distributions (black)}
\label{fig13.3.4}
\end{figure}

\begin{figure}[H]
\centering
\includegraphics[scale=1.00]{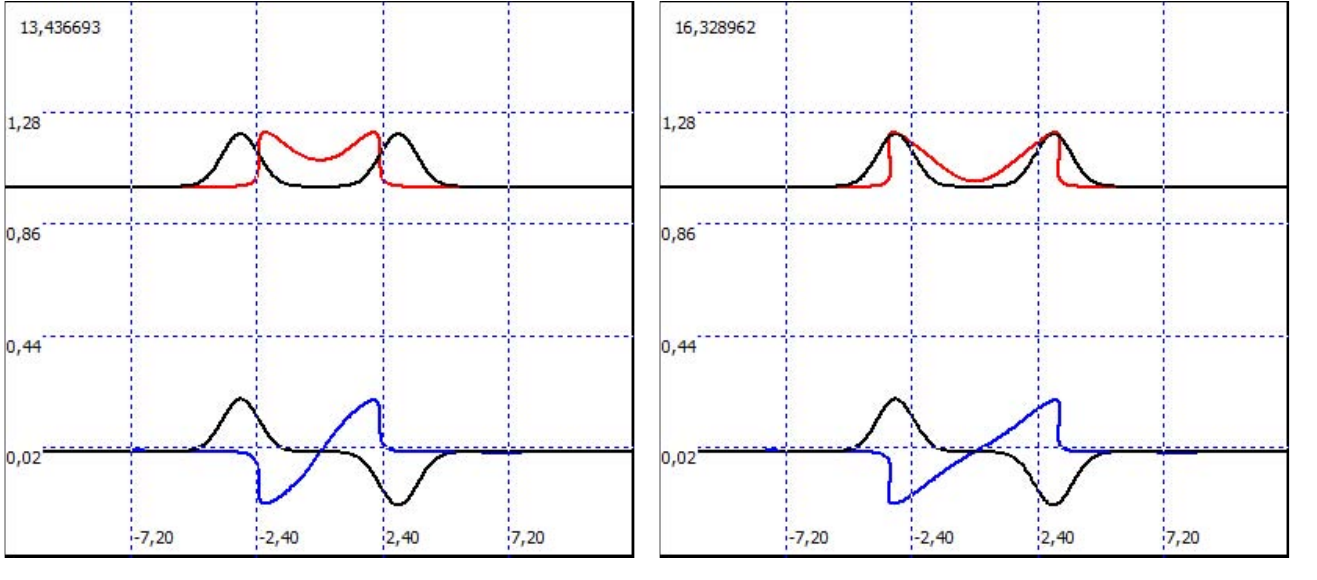}
\caption[Shallow water, waves interactions, $h=u^1$, $v=u^2$]{$h=u^1$ (red), $v=u^2$ (blue),  the initial distributions (black)}
\label{fig13.3.5}
\end{figure}

\begin{figure}[H]
\centering
\includegraphics[scale=1.00]{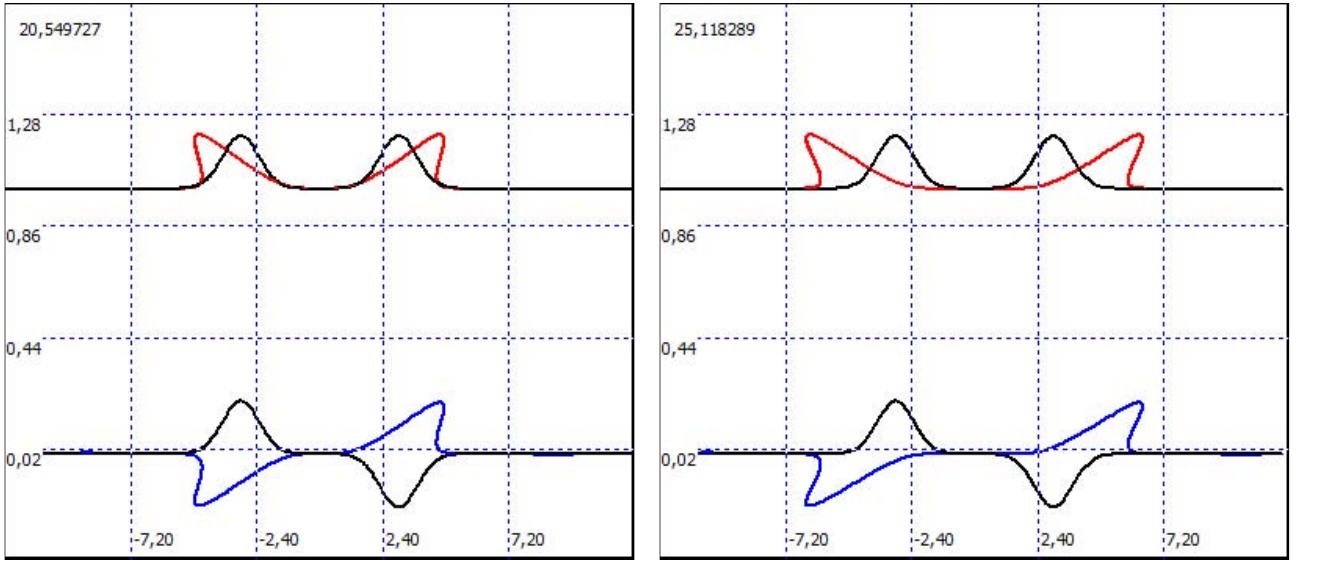}
\caption[Shallow water, waves interactions, $h=u^1$, $v=u^2$]{$h=u^1$ (red), $v=u^2$ (blue),  the initial distributions (black)}
\label{fig13.3.6}
\end{figure}

The position of the free surface and the distribution of the velocity field for different moments of time are shown on Fig.~\ref{fig13.3.1}--\ref{fig13.3.6}.

\subsection{Wave packet}\label{zhshArXiv:sec:5.B}

We take the perturbation of the free surface in the wave packet form, assuming that the velocity is equal to nought
\begin{equation}\label{zhshArXiv:eq:5.16}
u^1_0=1.0+0.1 \cos(3x) e^{x^2}, \quad
u^2_0=0.
\end{equation}

\begin{figure}[H]
\centering
\includegraphics[scale=1.00]{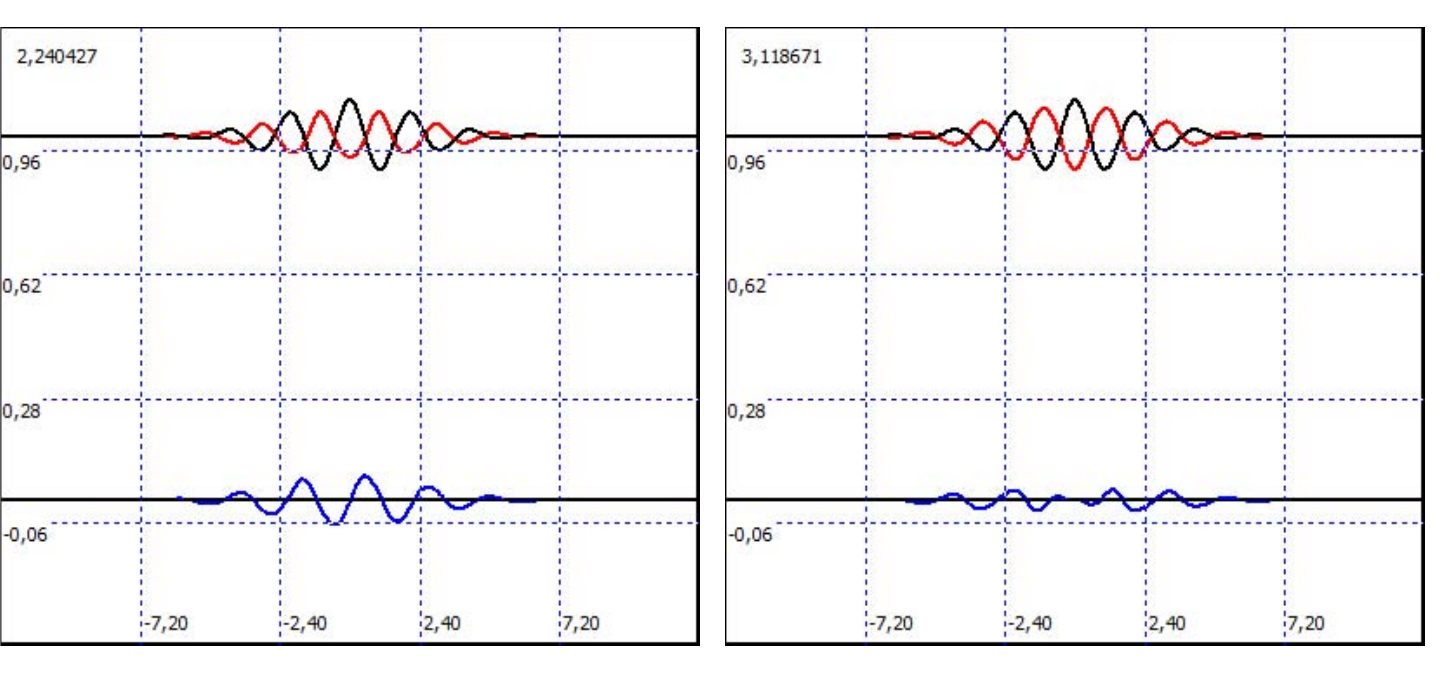}
\caption[Shallow water, wave packet, $h=u^1$, $v=u^2$]{$h=u^1$ (red), $v=u^2$ (blue),  the initial distributions (black)}
\label{fig13.4.1}
\end{figure}

\begin{figure}[H]
\centering
\includegraphics[scale=1.00]{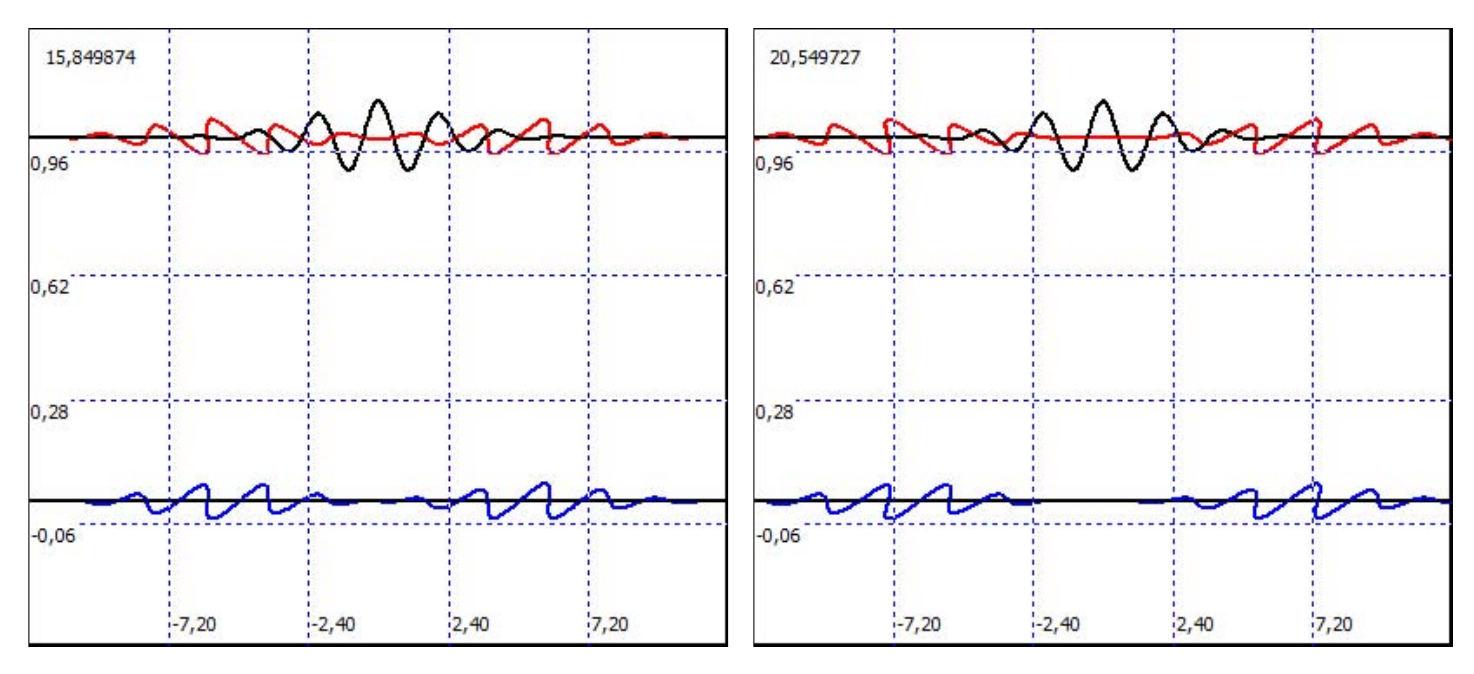}
\caption[Shallow water, wave packet, $h=u^1$, $v=u^2$]{$h=u^1$ (red), $v=u^2$ (blue),  the initial distributions (black)}
\label{fig13.4.2}
\end{figure}

The position of the free surface and the distribution of the velocity field for different moments of time are shown on Fig.~\ref{fig13.4.1}, \ref{fig13.4.2}.

\setcounter{equation}{0}

\section{Conclusions}

The choice to study equations of two-beam reduction of the dense soliton gas is not random selection.
First, these equations are degeneracy and, therefore, does not admit self-similar solutions.
Secondly, the results of the Sec.\,\ref{zhshArXiv:sec:4.D} show that the braking solution profile is impossible. All strong discontinuities are the so-called contact discontinuities, i.e. the discontinuities are moving along the characteristics. This, in particular, means that the proposed method allows to solve the Cauchy problem for arbitrary initial data, including discontinuous. Thirdly, the Riemann--Green function has a very simple form that allows us to easily analyze the solution and solve the Cauchy problem with initial data on an arbitrary curve. Note that the densities and fluxes of the conservation laws for equations of two-beam reduction of the dense soliton gas (as well as for equations of the zonal electrophoresis \cite{Zhuk_Shir_ArXiv_2014_2}) can be constructed as linear combinations of the original conservation laws (see (\ref{zhshArXiv:eq:4.09})--(\ref{zhshArXiv:eq:4.14})). Unfortunately, we could not use this method in the case of the shallow water equations.

As already mentioned, the numerical method is accurate, as it does not require any approximations of the original problem. The most efficient method operates  when there is an explicit expression for the Riemann--Green function. However, this method can be applied in cases when the Riemann--Green function is determined using the approximate solution of linear equations (\ref{zhshArXiv:eq:2.03})--(\ref{zhshArXiv:eq:2.07}), for example, in the form of an infinite series or by using numerical methods. Of course, in this case, the inevitably there are errors associated with the construction of the Riemann--Green function.

We say a few words about the Cauchy problem (\ref{zhshArXiv:eq:2.18})--(\ref{zhshArXiv:eq:2.20}). From our point of view, this problem is a generalization of the characteristics method to the case of two hyperbolic equations. Strictly speaking, formally, the method of characteristics for an arbitrary number of equations to construct is not very difficult. It is sufficient to consider the augmented system and construct the solution, for example, in the form of elementary waves \cite{RozhdestvenskiiYanenko,Bressan}. However, such equations are not closed. For the two equations the system can be closed, using the hodograph method and the Riemann--Green function. It would be interesting to build a similar scheme for solving the problem, bypassing the procedure to construct the Riemann--Green function of  (and possibly hodograph method), at least for two hyperbolic quasilinear equations.

\begin{acknowledgments}

The authors are grateful to N. M. Zhukova for proofreading the manuscript.
Funding statement. This research is partially supported by the Base Part of the Project no. 213.01-11/2014-1,
Ministry of Education and Science of the Russian Federation, Southern Federal University.

\end{acknowledgments}

\appendix

\section*{Appendix}\label{App:ZhS-0}

\renewcommand{\theequation}{A\arabic{section}.\arabic{equation}}%
\setcounter{equation}{0}

\section{Reduction of the Cauchy problem for two \\ hyperbolic quasilinear  PDE's to the Cauchy problem for ODE's}\label{zhshArXiv:sec:2}

Referring for details to \cite{Zhuk_Shir_ArXiv_2014_1,Zhuk_Shir_ArXiv_2014_2,SenashovYakhno}, here we give only a brief description of the method which allows to reduce
the Cauchy problem for two  hyperbolic  quasilinear PDE's  to  the Cauchy problem for ODE's.


\subsection{The Riemann invariants}\label{zhshArXiv:sec:2.A}

Let for a system of two hyperbolic PDE's, written in the Riemann invariants $R^1(x,t)$, $R^2(x,t)$,  we have the Cauchy problem at $t=t_0$
\begin{equation}\label{zhshArXiv:eq:2.01}
R^1_t+ \lambda^1(R^1,R^2)R^1_x=0, \quad R^2_t+ \lambda^2(R^1,R^2)R^2_x=0,
\end{equation}
\begin{equation}\label{zhshArXiv:eq:2.02}
R^1(\tau,t_0)=R^1_0(x), \quad R^2(x,t_0)=R^2_0(x),
\end{equation}
where $R^1_0(x)$, $R^2_0(x)$ are the functions determined on some interval of the axis $x$  (possibly infinite), $\lambda^1(R^1,R^2)$, $\lambda^2(R^1,R^2)$ are the given functions.

We recall that any system of two quasilinear equations can be reduce to the Riemann invariants (see \emph{e.g.} \cite{RozhdestvenskiiYanenko})

\subsection{Hodograph method}\label{zhshArXiv:sec:2.B}

Using the hodograph method for some conservation law
$\varphi_t+\psi_x=0$, where $\varphi(R^1,R^2)$ is the density, $\psi(R^1,R^2)$ is the flux, we write the equation \cite{SenashovYakhno}
\begin{equation}\label{zhshArXiv:eq:2.03}
\Phi_{R^1 R^2} + A(R^1,R^2)\Phi_{R^1} + B(R^1,R^2)\Phi_{R^2}=0,
\end{equation}
\begin{equation}\label{zhshArXiv:eq:2.04}
A(R^1,R^2)=\frac{\lambda^1_{R^2}}{\lambda^1-\lambda^2}, \quad
B(R^1,R^2)=-\frac{\lambda^2_{R^1}}{\lambda^1-\lambda^2}.
\end{equation}

\subsection{The Riemann--Green function}\label{zhshArXiv:sec:2.C}

Let the function $\Phi(R^1,R^2|r^1,r^2)$ is the Riemann--Green function  for equation (\ref{zhshArXiv:eq:2.03}).
The function $\Phi(R^1,R^2|r^1,r^2)$ of variables $R^1$, $R^2$ satisfies the given equation, and
the function $\Phi(R^1,R^2|r^1,r^2)$ of variables $r^1$, $r^2$ is the solution of the conjugate equation
\begin{equation}\label{zhshArXiv:eq:2.05}
\Phi_{r^1 r^2} - (A(r^1,r^2)\Phi)_{r^1} - (B(r^1,r^2)\Phi)_{r^2}=0,
\end{equation}
with conditions
\begin{equation}\label{zhshArXiv:eq:2.06}
(\Phi_{r^2} - A\Phi)\bigr|_{r^1=R^1}=0, \quad (\Phi_{r^1} - B\Phi)\bigr|_{r^2=R^2}=0,
\end{equation}
\begin{equation}\label{zhshArXiv:eq:2.07}
\Phi\bigr|_{r^1=R^1,r^2=R^2}=1.
\end{equation}
The methods of the Riemann--Green function construction are described, for example, in \cite{Copson,Chirkunov,Chirkunov_2,Courant,Ibragimov,SenashovYakhno}.

\subsection{Implicit solution of the problem}\label{zhshArXiv:sec:2.D}

It is convenient, to write the density of a conservation law, i.e. the function $\varphi(R^1,R^2)$,
in the form $\varphi(R^1,R^2|r^1,r^2)$
\begin{equation}\label{zhshArXiv:eq:2.08}
  \varphi(R^1,R^2|r^1,r^2) = M(r^1,r^2)\Phi(R^1,R^2|r^1,r^2), \quad   M(r^1,r^2)=\frac{2}{\lambda^2(r^1,r^2)-\lambda^1(r^1,r^2)}.
\end{equation}

The solution of (\ref{zhshArXiv:eq:2.01}), (\ref{zhshArXiv:eq:2.02}) can be represented in implicit form as \cite{SenashovYakhno}
\begin{equation}\label{zhshArXiv:eq:2.09}
R^1(x,t)=r^1(b)=R^1_0(b), \quad R^2(x,t)=r^2(a)=R^2_0(a),
\end{equation}
where $a$, $b$ are the new variables (Lagrange variables).

The connection between the new variables $a$, $b$ and old variables $x$, $t$ has the form
\begin{equation}\label{zhshArXiv:eq:2.10}
t=t(a,b), \quad x=x(a,b).
\end{equation}

Function $t=t(a,b)$ is calculated using the density of the conservation law $\varphi(R^1,R^2|r^1,r^2)$ and the initial data $R^1_0(x)$, $R^2_0(x)$ \cite{SenashovYakhno,Zhuk_Shir_ArXiv_2014_1,Zhuk_Shir_ArXiv_2014_2}
\begin{equation}\label{zhshArXiv:eq:2.11}
t(a,b)=t_0+\frac12\int\limits_{a}^{b}\varphi(R^1_0(\tau),R^2_0(\tau)|r^1(b),r^2(a))\,d\tau.
\end{equation}

Function $x=x(a,b)$ is calculated by analogously \cite{SenashovYakhno}, but this function is not required for further. We assume that this function is the given function.

If the equations (\ref{zhshArXiv:eq:2.10}) are solvable explicitly
\begin{equation}\label{zhshArXiv:eq:2.12}
  a=a(x,t), \quad b=b(x,t),
\end{equation}
then we have explicit solution
\begin{equation}\label{zhshArXiv:eq:2.13}
R^1(x,t)=R^1_0(b(x,t)), \quad R^2(x,t)=R^2_0(a(x,t)).
\end{equation}

\subsection{Solution on isochrones}\label{zhshArXiv:sec:2.E}

To construct the solution in the form (\ref{zhshArXiv:eq:2.09}) in \cite{Zhuk_Shir_ArXiv_2014_1} we proposed to solve the Cauchy problem for ODE's.

Fix some value $t=t_*$, specifying the level line (isochrone) of function $t(a,b)$
\begin{equation}\label{zhshArXiv:eq:2.14}
t_*=t(a, b).
\end{equation}
We assume that the isochrone is determined on the plane $(a,b)$  by the parametrical equations
\begin{equation}\label{zhshArXiv:eq:2.15}
a=a(\tau), \quad b=b(\tau),
\end{equation}
where $\tau$ is the parameter.

We choose the values $a_*$, $b_*$ which indicate some point on isochrone $t=t_*$
\begin{equation}\label{zhshArXiv:eq:2.16}
t_*=t(a_*, b_*).
\end{equation}
In practice, the values of $a_*$, $b_*$ one can choose using the  line levels of function $t(a,b)$ for some ranges of parameters $a$, $b$.

The coordinate $x$ on isochrone, obviously, is determined by the expression
\begin{equation}\label{zhshArXiv:eq:2.17}
x=x(a(\tau), b(\tau))\equiv X(\tau).
\end{equation}

To determine the $a(\tau)$, $b(\tau)$, $X(\tau)$ we have the  Cauchy problem \cite{Zhuk_Shir_ArXiv_2014_1,Zhuk_Shir_ArXiv_2014_2}
\begin{equation}\label{zhshArXiv:eq:2.18}
\frac{da}{d\tau}=-t_b(a,b), \quad
\frac{db}{d\tau}=t_a(a,b),
\end{equation}
\begin{equation}\label{zhshArXiv:eq:2.19}
\frac{dX}{d\tau}=(\lambda^2(r^1(b),r^2(a))-\lambda^1(r^1(b),r^2(a)))t_a(a,b) t_b(a,b),
\end{equation}
\begin{equation}\label{zhshArXiv:eq:2.20}
a\bigr|_{\tau=0}=a_*, \quad b\bigr|_{\tau=0}=b_*, \quad
X\bigr|_{\tau=0}=X_*.
\end{equation}
Here the values $a_*$, $b_*$ are given. To determine  $X_*$ we need to solve the problem
\begin{equation}\label{zhshArXiv:eq:2.21}
\frac{dY(b)}{db}=x_b(a_*,b)=\lambda^2(r^1(b),r^2(a_*))t_b(a_*,b), \quad Y(a_*)=a_*.
\end{equation}
Integrating from $a_*$ to $b_*$ we get
\begin{equation}\label{zhshArXiv:eq:2.22}
X_*=Y(b_*).
\end{equation}
Note, that $X_*=x(a_*,b_*)$ is the $x$ coordinate corresponding to $\tau=0$.

Solving (\ref{zhshArXiv:eq:2.18})--(\ref{zhshArXiv:eq:2.20})
we obtain the solution on isochrone
\begin{equation}\label{zhshArXiv:eq:2.23}
R^1(x,t_*)=R^1_0(b(\tau)), \quad R^2(x,t_*)=R^2_0(a(\tau)), \quad
x=X(\tau).
\end{equation}
Changing the parameter $\tau$ we obtain the solution which depends on  $x$ as the fixed $t=t_*$. Pay  attention to the
fact that the right hand sides of differential equations, in particular, $t_a(a,b)$, $t_b(a,b)$ are easily computed with help of (\ref{zhshArXiv:eq:2.08}), (\ref{zhshArXiv:eq:2.09}), (\ref{zhshArXiv:eq:2.11}).

\renewcommand{\theequation}{B\arabic{section}.\arabic{equation}}%
\setcounter{equation}{0}

\section{Additional simplification}\label{zhshArXiv:sec:AppB}

Using another parameters $\theta$ instead $\tau$
\begin{equation}\label{zhshArXiv:eq:AppB.01}
d\tau = {(R^1_0(a)-R^2_0(a))(R^1_0(b)-R^2_0(b))}d\theta
\end{equation}
one can simplify the equations (\ref{zhshArXiv:eq:4.37}), (\ref{zhshArXiv:eq:4.38}).

In this case instead of the equations (\ref{zhshArXiv:eq:4.37}), (\ref{zhshArXiv:eq:4.38}) we get
\begin{equation}\label{zhshArXiv:eq:AppB.02}
\frac{da}{d\theta}=R^2_0(a)-R^1_0(a), \quad
\frac{db}{d\theta}=R^2_0(b)-R^1_0(b), \quad
\frac{dX}{d\theta}=R^2_0(a)-R^1_0(b).
\end{equation}

\end{document}